\documentclass[aps,nofootinbib,showpacs,preprintnumbers,twocolumn,superscriptaddress,prl]{revtex4}

\usepackage{amsmath, amsthm, amssymb}
\usepackage{graphicx}
\usepackage{bm}
\usepackage{epsfig}
\usepackage[T1]{fontenc}
\usepackage[latin9]{inputenc}
\usepackage{color}
\usepackage{hyperref}
\usepackage{subfigure}

\renewcommand{\vec}[1]{\mathbf{#1}}

\begin{document}

\title{Unconventional superconductivity on the triangular lattice Hubbard model}

\author{Kuang Shing Chen}
\email{kchen5@lsu.edu}
\affiliation{Department of Physics and Astronomy, Louisiana State University, Baton Rouge, LA 70803, USA}
\author{Zi Yang Meng}
\email{zmeng@lsu.edu}
\affiliation{Department of Physics and Astronomy, Louisiana State University, Baton Rouge, LA 70803, USA}
\affiliation{Center for Computation and Technology, Louisiana State University, Baton Rouge, LA 70803, USA}
\author{Unjong Yu}
\affiliation{GIST-College, Gwangju Institute of Science and Technology, Gwangju 500-712, Korea}
\author{Shuxiang Yang}
\affiliation{Department of Physics and Astronomy, Louisiana State University, Baton Rouge, LA 70803, USA}
\affiliation{Center for Computation and Technology, Louisiana State University, Baton Rouge, LA 70803, USA}
\author{Mark Jarrell}
\affiliation{Department of Physics and Astronomy, Louisiana State University, Baton Rouge, LA 70803, USA}
\affiliation{Center for Computation and Technology, Louisiana State University, Baton Rouge, LA 70803, USA}
\author{Juana Moreno}
\affiliation{Department of Physics and Astronomy, Louisiana State University, Baton Rouge, LA 70803, USA}
\affiliation{Center for Computation and Technology, Louisiana State University, Baton Rouge, LA 70803, USA}
\begin{abstract}
Using large-scale dynamical cluster quantum Monte Carlo simulations, 
we explore the unconventional superconductivity in the hole-doped Hubbard model 
on the triangular lattice. Due to the interplay of 
electronic correlations, geometric frustration, and Fermi surface topology, 
we find a doubly degenerate singlet pairing state at an interaction strength close to the bare bandwidth. 
Such an unconventional superconducting state is mediated by antiferromagnetic spin fluctuations
along the $\Gamma$-$K$ direction, where the Fermi surface is nested. 
An exact decomposition of the irreducible particle-particle vertex further confirms 
the dominant component of the effective pairing interaction comes from the spin channel. 
Our findings suggest the existence of chiral $d +i d$ superconductivity in hole-doped Hubbard 
triangular lattice in strongly correlated regime, and provide insight to the superconducting phases 
of the water-intercalated sodium cobaltates Na$_{x}$CoO$_{2} \cdot y$H$_{2}$O, as well as the 
organic compounds $\kappa$-(ET)$_{2}$X and Pd(dmit)$_{2}$.

\end{abstract}


\maketitle

\paragraph*{Introduction.-}
Since the discovery of the Cu-based high temperature superconductors, the search for new 
unconventional superconductors is among the central topics in condensed-matter 
physics \cite{Sigrist91,McKenzie97}.  The water-intercalated sodium cobaltates 
Na$_{x}$CoO$_{2} \cdot y$H$_{2}$O \cite{Takada03, Schaak03, Foo04} and two families of 
organic charge-transfer salts $\kappa$-(ET)$_{2}$X and 
Pd(dmit)$_{2}$ \cite{Shimizu03,Kurosaki05,SYamashita08,MYamashita08,Itou08,SYamashita11} 
are of particular interest.  The underlying structure of these layered materials is 
the geometrically frustrated triangular lattice. The competition between electronic 
correlations and geometric frustration yields novel phenomena \cite{Liu05,Kyung06,Powell07}. For example, the most 
frustrated members of the $\kappa$-(ET)$_{2}$X and Pd(dmit)$_{2}$ families are 
believed to host quantum spin liquid states \cite{Anderson73,Kyung06,Powell07}, 
and the recently discovered 5K superconducting phase in Na$_{x}$CoO$_{2}.y$H$_{2}$O might 
be a chiral state which breaks parity and time reversal giving rise to interesting edge 
modes that can carry quantized particle and spin currents \cite{Ogata03,Watanabe04,Braunecker05,Powell07,Zhou08,Nandkishore12,Kiesel13}.

The layered triangular lattice compound Na$_{x}$CoO$_{2} \cdot y$H$_{2}$O has a superconducting 
dome for $x\sim0.3$, $y\sim1.3$ at $T_c\sim5$ K~\cite{Takada03, Schaak03, Foo04}. 
Due to intercalation, its electronic structure is effectively two-dimensional.
A very rich phase diagram has been mapped out for a range of Na concentrations~\cite{Foo04}; 
however, the nature of the superconducting phase has remained poorly understood. 
Recent measurements on high quality single crystals \cite{Zheng06a} show that the 
spin contributions to the Knight shift decreases below $T_c$ along the $a$ and $c$ axes, 
supporting the notion that the Cooper pairs are formed in a spin-singlet state. The temperature and doping 
dependence of the Knight shift and the relaxation rate above $T_c$ provide evidence of antiferromagnetic 
correlations \cite{Zheng06a,Zheng06b}. 

There are a number of theoretical proposals for the unconventional superconductivity in the cobaltates. 
The underlying triangular lattice 
allows a doubly degenerate $E_2$ representation of the superconducting order parameter 
with $d_{x^2-y^2}$ and $d_{xy}$ degenerate states~\cite{Sigrist91,Powell07,Nandkishore12,Kiesel12}, 
raising the exciting possibility of a time-reversal symmetry breaking 
chiral $d_{x^2-y^2}\pm id_{xy}$ superconductor~\cite{Kumar03,Ogata03,Watanabe04}. Earlier studies of the cobaltates draw 
analogy to the cuprates and  employed either phenomenological RVB mean field theory~\cite{Kumar03,Baskaran03}, 
or slave boson mean-field approach \cite{Lee04,Braunecker05} to provide signatures of a spin-singlet 
$d+id$ pairing state. Also there are variational mean-field theory~\cite{Ogata03} and variational
Monte Carlo studies~\cite{Watanabe04,Liu05}. Recent studies  
of the sodium cobaltates using a
Gutzwiller projection supplemented by symmetry arguments~\cite{Zhou08}, RVB mean field theory~\cite{Powell07}, 
as well as the multi-orbital 
functional renormalization group \cite{Kiesel13}, reveal a rich phase diagram with an anisotropic $d+id$ 
phase and a possible topological quantum phase 
transition through a nodal superconducting state. However, 
prior approaches suffer either from their mean-field nature, or their incapability of capturing 
correlation effects in the strong coupling regime.  Hence, there is an urgent need of unbiased studies, 
where the interplay of strong electronic correlations and geometric frustration can be treated in a 
non-perturbed fashion.

The simplest model that captures the essential physics of the cobaltates 
is the single-band Hubbard model on a triangular lattice. In this Rapid Communication, we explore the low-energy 
properties of this model by large-scale dynamical cluster quantum Monte Carlo simulations \cite{Maier05}.  
We focus on the different superconducting instabilities in the hole-doped side of the phase 
diagram.  To the best of our knowledge, this is the first study of the hole-doped Hubbard model on the 
triangular lattice exploring the 
pairing symmetries on different cluster sizes. 
Clusters up to size $N_c=12$ allow  a greater momentum resolution and higher quality data on 
the spectral function, self energy, and different superconducting susceptibilities. Therefore, we obtain an
unambiguous signature of an unconventional doubly-degenerate superconducting state in the strong to intermediate
coupling region. By explicitly comparing the pairing susceptibility in the $s$-, $d_{x^2-y^2}$-, $d_{xy}$-wave 
singlet channels and the $f$-wave triplet channel,  we find that the $d_{x^2-y^2}$ 
and $d_{xy}$ components are most divergent and extrapolate to the same $T_c$ within our numerical 
accuracy.  We identify that the pairing is mediated by strong spin fluctuations along the 
antiferromagnetically (AF) ordered wavevector on the $\Gamma$ to $K$ direction. The Fermi 
surface (FS) is nested along this AF wavevector, but the system only orders 
at half-filling in the Heisenberg limit. 
An exact decomposition of the irreducible particle-particle vertex furthermore reveals the dominant
part in the effective pairing interaction comes from the spin channel.
 
\paragraph*{Formalism.-}
The Hamiltonian of the system is 
$H=\sum_{\vec{k}\sigma}(\epsilon_{\vec{k}}^{0}-\mu)c_{{\vec{k}}\sigma}^{\dagger}c_{{\vec{k}}\sigma}^{\phantom{\dagger}}+U\sum_{i}n_{{i}\uparrow}n_{{i}\downarrow}$, 
where $c_{{\vec{k}}\sigma}^{\dagger}(c_{{\vec{k}}\sigma})$ is the creation (annihilation) operator for 
electrons with momentum ${\vec{k}}$ and spin $\sigma$, $\mu$ is the chemical potential, $n_{i\sigma} =c_{i\sigma}^{\dagger}c_{i\sigma}$ 
is the number operator, and the bare dispersion is given by 
$\epsilon_{\vec{k}}^{0}=-2t\cos(k_{x})-4t\cos( \sqrt{3}k_{y}/2)\cos(k_{x}/2)$ with $t$ being the hopping 
amplitude between nearest neighbor sites, and $U$ the on-site Coulomb repulsion.

We investigate one- and two-particle properties of the model using  the dynamical cluster 
approximation (DCA)~\cite{Hettler98} with weak-coupling continuous time quantum 
Monte Carlo (CTQMC)~\cite{Rubtsov05} as the cluster solver. The DCA maps the original 
lattice onto a periodic cluster of size $N_c$ embedded in a self-consistently determined 
host. Spatial correlations inside a cluster are treated explicitly while those at longer 
length scales are described at the mean-field level.  In this work we choose clusters of 
sizes $N_c=4,6,8$ and $12$. We study inverse temperatures up to $\beta t=16.5$.  We obtain 
the cluster self-energy $\Sigma(\vec{K},\omega)$ via the maximum entropy 
method~\cite{Jarrell96} (MEM) applied directly to the Matsubara-frequency self energies 
calculated by the DCA-CTQMC~\cite{Wang09,Chen12}.  We then interpolate the 
$\Sigma(\vec{K},\omega)$ to obtain the lattice self energy, $\Sigma(\vec{k},\omega)$, 
and lattice spectral function, $A(\vec{k},\omega)$.  

To obtain various susceptibilities, $\chi(T)$, we extract the irreducible vertex 
function $\Gamma$ via the Bethe-Salpeter equation from the two-particle Green 
function measured on the cluster, then employing 
$\chi(T)=\displaystyle \frac{\chi_0}{1-\Gamma\chi_0}$, 
where $\chi_0$ is the bare susceptibility constructed from the dressed one-particle 
lattice Green function.  The superconducting pairing susceptibilities are obtained 
from the particle-particle channel, and the charge and spin susceptibilities are 
obtained from the particle-hole channel. We further separate the pairing 
susceptibilities explicitly into spin singlet and triplet channels, where in the 
singlet channel we project the $\chi(T)_{pairing}$ onto $s$-, $d_{x^2-y^2}$-, and 
$d_{xy}$-wave, and in the triplet channel we project it onto the $f$-wave channel, 
with the corresponding form factors~\cite{Sigrist91, Kiesel13}.

To explore the pairing mechanism we decompose the particle-particle pairing vertex 
$\Gamma$ into the fully irreducible vertex $\Lambda$, the charge ($S=0$) particle-hole 
contribution, $\Phi_{c}$, and the spin ($S=1$) particle-hole contribution, $\Phi_s$, 
through the parquet equation, $\Gamma=\Lambda+\Phi_c+\Phi_s$~\cite{Maier06}. 
We furthermore project the previous expression using different form factors such as 
$d_{x^2-y^2}$ and $d_{xy}$,
\begin{equation}
V_{d_{x^2-y^2}/d_{xy}}=V^{\Lambda}_{d_{x^2-y^2}/d_{xy}}+V^{C}_{d_{x^2-y^2}/d_{xy}}+V^{S}_{d_{x^2-y^2}/d_{xy}},
\end{equation} 
where each term is the projected component of the corresponding term in the parquet 
equation~\cite{Yang11}.  In this way, we are able to distinguish which component 
contributes the most to the effective pairing interaction. One important point to note is that 
as we have controlled information about the two particle vertex function in momentum and frequency, we 
don't need assume any kind of pairing mechanism a prior, but can numerically prove which channel is dominant
in the pairing interaction. This is a qualitatively improvement than many weak-coupling approaches where 
one channel (usually spin) is always assumed to dominate the pairing interaction~\cite{Schmalian98,Kino98,Kondo98,Kontani03}.

\begin{figure}[tp]
\centering
  \includegraphics[width=\columnwidth]{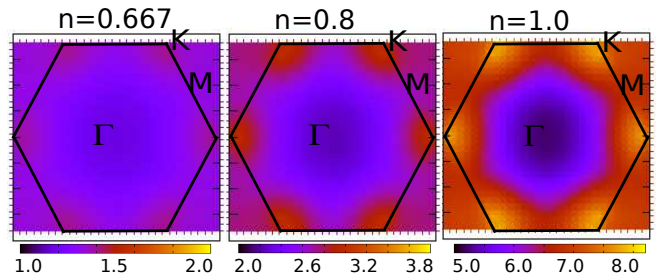}
  \caption{(Color online) Cluster spin susceptibility for $N_c=12$, interaction strength $U=8.5t$, 
temperature $T=0.1t$ and different fillings, $n=0.667$, $0.8$, and $1$.
  \label{fig:ChiQ}
}
\end{figure}

\paragraph*{Results.-}
Fig.~\ref{fig:ChiQ} displays the cluster spin susceptibility at different fillings, $n=0.667$, $0.8$, and $1$, 
and coupling $U=8.5t$. The data are obtained from DCA-CTQMC simulations with cluster size $N_c=12$, and we 
interpolate the cluster susceptibility into the entire Brillouin zone (BZ). 
At very high hole-doping, $n=0.667$, the susceptibility is mostly flat.  As the filling 
increases, $n=0.8$, the spin susceptibility develops six bumps at the $K$ points. 
When $n=1$, the bumps become more pronounced.  The vector connecting $\Gamma$ to 
$K$ is the antiferromagnetically
ordered wave vector ($\vec{Q_{AF}}$) in the Heisenberg limit of the half-filled model. 
The cluster spin susceptibility demonstrates that the antiferromagnetic fluctuations become 
stronger as the filling moves towards $n=1$.  The pairing of electrons may be mediated 
by these fluctuations \cite{Miyake86,Scalapino86,Schmalian98,Kino98,Kondo98,Kontani03}.

\begin{figure}[tp]
\centering
  \includegraphics[width=\columnwidth]{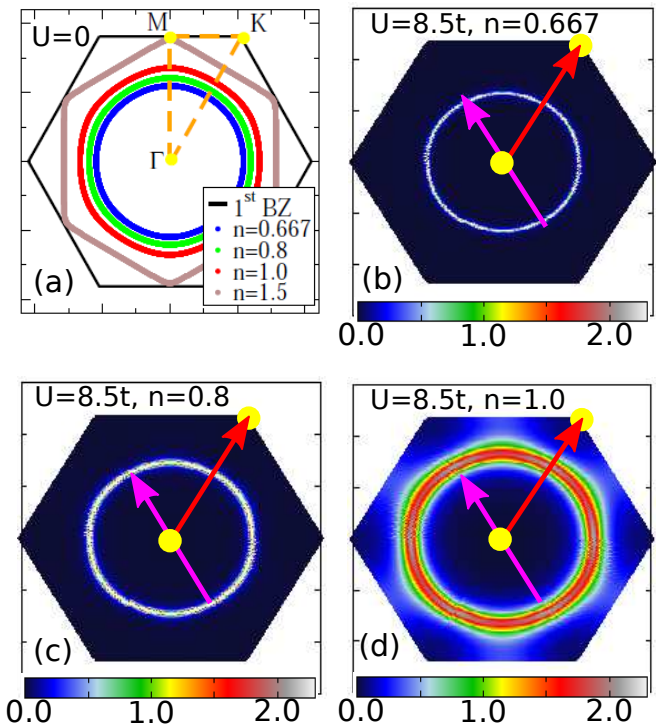}
  \caption{(Color online) (a) First Brillouin zone, the symmetric path $\Gamma-M-K-\Gamma$ and the non-interacting 
Fermi surface at different band fillings. (b), (c), (d) Spectral function $A(\vec{k}, \omega=0)$ on the Fermi 
surface for $N_c=12$ DCA-CTQMC simulations with  $U=8.5t$,  $T=0.1t$, 
and $n=0.667$ in (b), $n=0.8$ in (c), 
and $n=1$ in (d). Red arrow is the AF ordered wave vector ($\vec{Q_{AF}}$) and the pink arrow is after 
shifting its center to $\Gamma$ and rotating it by 60$^o$.
  \label{fig:FS}}
\end{figure}

Fig.~\ref{fig:FS} shows the Fermi surface (FS) at the same fillings used in the previous figure.
Fig.~\ref{fig:FS} (a) corresponds to the non-interacting limit. At $n=0.667$, $0.8$ and $1$ 
the FS is close to a perfect circle. The van Hove singularity in the non-interacting band 
structure is present at $n=1.5$ with saddle points at $M$. 
One-loop RG calculations~\cite{Honerkamp03} show the FS in the hole-doped side is stable against 
weak Coulomb interactions. However, under strong interaction, the FS begins to deform. 
Fig.~\ref{fig:FS} (b) displays the FS at $n=0.667$, $U=8.5t$, which is slightly deformed towards a hexagon. 
The red arrow corresponds to $\vec{Q_{AF}}$, while the pink arrow is this vector shifting its center to $\Gamma$ 
and rotating it by 60$^o$.
For $n=0.667$ the pink arrow is longer than the diameter of the FS so there is no nesting effect, and 
we do not observe superconductivity at this filling. 
In Fig.~\ref{fig:FS} (c), $n=0.8$, the FS is more  deformed, the $\vec{Q_{AF}}$ 
now connects significant sections of the FS, and as illustrated in Fig.~\ref{fig:ChiQ}, 
the AF fluctuations are stronger. The nesting effect and the strong AF fluctuations 
together give rise to diverging pairing susceptibilities  
at filling $n=0.8$ and $0.9$, as discussed below. 
At half-filling, $n=1$, the FS is further deformed towards a hexagon, 
but the spectral weight become less coherent. 
Interestingly, the nesting vector now is shorter than the diameter of the FS. Hence, even though the 
AF fluctuations are the strongest here, electrons on the FS are hard to pair by $\vec{Q_{AF}}$, 
the system is rather subject to a Mott transition, whose novel features are beyond the scope of this paper~\cite{Liu05,Kyung06,Powell07}.

\begin{figure}[tp]
\centering
  \includegraphics[width=\columnwidth]{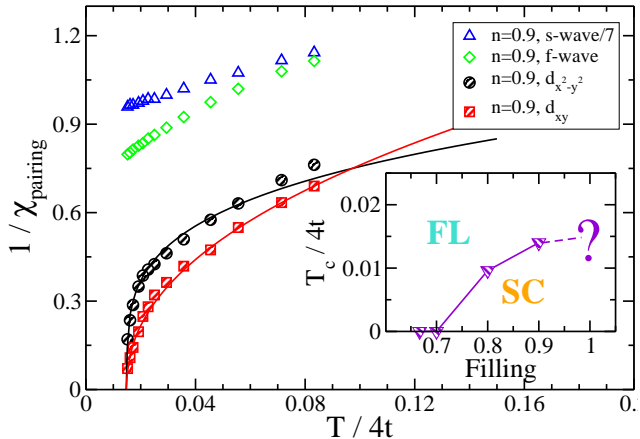}
  \caption{(Color online) Inverse pairing susceptibility, $1/\chi_{pairing}$, for $N_c=6$, $U=8.5t$ and 
$n=0.9$. The singlet $s$-wave and triplet $f$-wave do not diverge, whereas the singlet pairing channels 
with $d_{x^2-y^2}$- and $d_{xy}$-wave symmetry show a divergencency at the same $T_c$. Note that we have 
multiplied by a factor of $7$ the $s$-wave pairing susceptibility in order to use the same vertical scale. 
Inset, the superconducting transition temperature $T_c$ as a function of doping. FL and SC label the 
Fermi liquid and superconducting regions, respectively. The nature of system is still unclear at half-filling, 
hence a question mark in the inset.
  \label{fig:sdfwaves}
}
\end{figure}

Fig.~\ref{fig:sdfwaves} displays the inverse pairing susceptibility as a function of
temperature, $1/\chi_{pairing}(T)$, at filling $n=0.9$, $U=8.5t$, and $N_c=6$. 
Here we explicitly project the lattice pairing susceptibility in the $s$-, $d_{x^2-y^2}$-, 
and $d_{xy}$-wave singlet channels and the $f$-wave triplet channel
by using the appropriate form factors. Fig.~\ref{fig:sdfwaves} shows that 
the two singlet $d$-wave components are the most divergent ones. 
Within our numerical resolution their $1/\chi_{pairing}$ extrapolate to zero at the  
same superconducting transition temperature, $T_c$. This implies that the superconducting order parameter
is doubly degenerate with components $d_{x^2-y^2}$ and $d_{xy}$. Based on symmetry arguments any linear 
combination of both d-wave components is possible below $T_c$. However, both Ginzburg-Landau and BCS-type 
mean-field approaches favor superconducting phases that break the time-reversal symmetry for singlet multicomponent 
superconductors~\cite{Joynt02,Kuznetsova05},  such as the $d+id$ singlet pairing state predicted in graphene~\cite{Nandkishore12, Kiesel12} and
the cobaltates~\cite{Zhou08,Kiesel13}. Therefore, our findings support the possibility of a chiral  $d+id$
superconducting phase in the hole-doped triangular Hubbard model. 
The inset of Fig.~\ref{fig:sdfwaves} shows the phase diagram for different doping 
concentrations based on $N_c=6$ DCA-CTQMC simulations. $T_c$ becomes finite for doping larger than
$n=0.7$ due to the onset of FS nesting and strong AF correlations, and 
increases as $n$ aproaches $1$, reflecting that the AF fluctuations become stronger towards 
half-filling. However, the nature of the ground state at half-filling is still 
unclear due to a worsened minus-sign problem in our simulations, hence we put a question mark
in the inset, and are investigating this case at the moment.

\begin{figure}[tp]
\centering
   \includegraphics[width=\columnwidth]{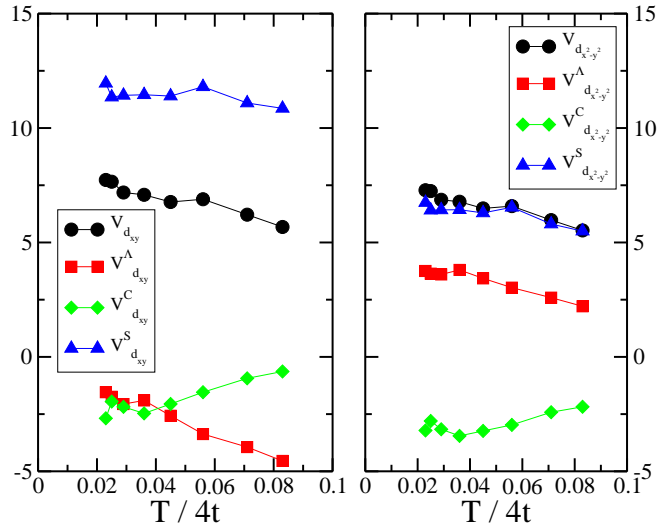}
  \caption{(Color online) Left: $d_{xy}$ projected contributions to the pairing vertex 
$V_{d_{xy}}$, from the fully irreducible vertex $V^{\Lambda}_{d_{xy}}$, charge $V^{C}_{d_{xy}}$ 
and spin $V^{S}_{d_{xy}}$ cross channels versus $T$ at $n=0.9$,  $U=8.5t$. 
Right: the $d_{x^2-y^2}$-wave projection of the same quantities. In both cases, the contribution 
to the pairing interaction from spin channel is clearly dominant. 
  \label{fig:vertex}
}
\end{figure}

To shine light on  the dominant contribution to the pairing interaction, we use the parquet equations 
to decompose the irreducible particle-particle vertex function, and project each term onto its 
$d_{x^2-y^2}$ and $d_{xy}$ components. 
The results are presented in Fig.~\ref{fig:vertex} for a DCA-CTQMC simulation with cluster 
size $N_c=12$, $U=8.5t$ and filling $n=0.9$. The left, right panels correspond to the $d_{xy}$, 
$d_{x^2-y^2}$ projection of the parquet equations, respectively. In both cases, 
the dominant contribution to the effective pairing interactions $V_{d_{xy}}$ 
and $V_{d_{x^2-y^2}}$ is from the magnetic, spin $S=1$, particle-hole channel, 
$V^S_{d_{xy}}$ and $V^S_{d_{x^2-y^2}}$.  In fact, we also find that the 
pairing interaction, $V_{d_{xy}/d_{x^2-y^2}}(\vec{k}-\vec{k'})$ is peaked at 
momentum transfer $|\vec{k}-\vec{k'}|=|\vec{Q_{AF}}|$. The vertex decomposition confirms that 
this peak comes from the spin channel $V^S_{d_{xy}/d_{x^2-y^2}}(\vec{Q_{AF}})$ (not shown). 
Note that both $N_c=6$ and $12$ size clusters have the cluster points connected by $\vec{Q_{AF}}$. 
From the BCS gap equation~\cite{Joynt02}
\begin{equation}
\Delta_{\vec{k}} = -\frac{1}{N} \sum_{\vec{k'}} V^{SC}(\vec{k}-\vec{k'})\frac{\Delta_{\vec{k'}}}{2E(\vec{k'})}\tanh(\frac{E(\vec{k'})}{2T}),
\end{equation}
where $E(\vec{k})=\sqrt{\epsilon_{\vec{k}}^2 + \Delta_{\vec{k}}^2  }$, we infer that if the 
superconducting pairing interaction $V^{SC}(\vec{k}-\vec{k'})$ is peaked 
at $\vec{Q_{AF}}$, the order parameters $\Delta_{\vec{k}}$ which correspond to $d_{x^2-y^2}$, 
$d_{xy}$ and $f$-waves are equally favored in the $N_c=6$ and $12$ clusters.  Our results 
suggest that $d_{xy}$ and $d_{x^2-y^2}$ singlet pairing are favored over the $f$-wave 
triplet pairing, probably because $f$-wave has a more complex nodal structure than 
the two $d$-waves~\cite{Sigrist91}.

\paragraph*{Conclusion.-}
Using large-scale dynamical cluster quantum Monte Carlo simulations, 
we find a doubly degenerate singlet pairing state at interaction strength close 
to the bare bandwidth and filling larger than $n=0.7$ in the hole-doped 
Hubbard model on  the triangular lattice. Our findings suppport the  
presence of a chiral  $d+id$ singlet superconducting phase in this model.
The pairing mechanism comes from antiferromagnetic 
spin fluctuations at the magnetic order wavevector nesting the deformed FS. A decomposition 
of the vertex further confirms that the spin channel contributes the most to the effective 
pairing interaction.

\begin{acknowledgments}
We acknowledge 
G.\ Chen, 
K.\ Kanoda,
K.-M.\ Tam, 
and 
Y.\ Zhou
for useful discussions.  
This work is supported by NSF OISE-0952300 (KSC, JM), and the EPSCoR Cooperative 
Agreement EPS-1003897 (ZYM). 
Additional support was provided by the DOE SciDAC grant DE-FC02-10ER25916 (MJ, SXY)
and the Korean National Research Foundation Grant NRF-2011-0013866 (UY). 
Supercomputer support was provided by the NSF XSEDE grant number DMR100007, the Louisiana Optical 
Network Initiative, and HPC@LSU computing resources.
\end{acknowledgments}

\end{document}